\documentclass[runningheads]{llncs}
\usepackage{graphicx}
\usepackage{subfigure}
\usepackage{booktabs} 
\usepackage{amsmath}
\usepackage{mathtools}
\usepackage{amssymb,amsfonts}
\usepackage{hyperref}
\usepackage{authblk}
\usepackage{algorithm}
\usepackage{algorithmic}

\usepackage{xcolor}
\usepackage{array}
\usepackage{bbding}
\newcolumntype{C}[1]{>{\centering}p{#1}}
\usepackage{hyperref}[colorlinks,linkcolor=black]
\usepackage{breakurl}[hyphenbreaks]
\definecolor{Red}{rgb}{1,0,0}
\begin{document}
\title{
%Generative Adversarial Networks for
Solving Low-Dose CT Reconstruction via GAN with Local Coherence}
% \title{Low-Dose Computed Tomography Reconstruction with Slices Coherence }

\author{Wenjie Liu\inst{1}\orcidID{0000-0002-4524-8507} }
%	\and
%Hu Ding\inst{1(}\Envelope\inst{)}\orcidID{0000-0002-1307-6077}}
% index{Wenjie, Liu}
% index{Hu, Ding}

\authorrunning{Liu }
%and Hu
% \institute{Anonymous Organization}
\institute{School of Computer Science and Technology, \\
University of Science and Technology of China, Hefei, Anhui, China, \\
\email{lwj1217@mail.ustc.edu.cn}\\
%,huding@ustc.edu.cn
}

\maketitle              % typeset the header of the contribution
\begin{abstract}
The Computed Tomography (CT) for diagnosis of lesions in human internal organs is one of the most fundamental topics in medical imaging. Low-dose CT, which offers reduced radiation exposure, is preferred over standard-dose CT, and therefore its reconstruction approaches have been extensively studied. However, current low-dose CT reconstruction techniques mainly rely on model-based methods or deep-learning-based techniques, which often ignore the coherence and smoothness for sequential CT slices. To address this issue, we propose a novel approach using generative adversarial networks (GANs)  with enhanced local coherence. The proposed method can capture the local coherence of adjacent images by optical flow, which yields significant improvements in the precision and stability of the constructed images. We evaluate our proposed method on real datasets and  the experimental results suggest  that it can outperform existing state-of-the-art reconstruction approaches significantly. 

\end{abstract}
%
%local coherence
%
\section{Introduction} \label{introduction}
Computed Tomography (CT) is one of the most widely used  technologies in medical imaging, which can assist doctors for diagnosing the lesions in human internal organs. 
Due to harmful radiation exposure of standard-dose CT, the low  dose  CT is more preferable in clinical application~\cite{chen2017low,ding2021deep,wolterink2017generative}. However, when the dose is low together with   the issues like sparse-view or limited angles, it becomes quite challenging to reconstruct high-quality CT images.
%from the signal received by the equipment. 
%Most of the existing researches define
%%%%%%%%%%%%%%%%%%%%%%%%%
The  high-quality CT images are important to improve the performance of diagnosis in clinic~\cite{sori2021dfd}. 
%In addition, the denoised CT images have demonstrated their potential to facilitate accurate diagnosis by doctors and improve the performance of recognition tasks for deep learning
%In addition, the denoised CT images are more conducive to doctors in diagnosis and the recognition tasks in deep learning
%%%%%%%%%%%%%%%%%%%%%%%%%%%%%%%
In mathematics, we model the CT imaging as the following procedure
%The problem of CT image reconstruction is often modeled  as the following inverse problem 
\begin{equation} \label{for-inverse}
    \mathbf{y}=\mathcal{T}(\mathbf{x^r})+\delta, 
\end{equation}
where $\mathbf{x^r} \in \mathbb{R}^d$ denotes the \textbf{unknown} ground-truth picture,  $\mathbf{y} \in \mathbb{R}^m$ denotes the received measurement, and $\delta$ is the noise. The function $\mathcal{T}$ represents the forward operator that is analogous to the Radon transform, which is widely used in medical imaging~\cite{ramm2020radon,toft1996radon}. The problem of CT reconstruction is to recover $\mathbf{x^r}$ from the received $\mathbf{y}$.
%, and noise is  represented by  $\delta$.

Solving the inverse problem of (\ref{for-inverse}) is often very challenging if there is no any  additional information.
If the forward operator $\mathcal{T}$ is well-posed and $\delta$ is neglectable, we know that an approximate $\mathbf{x^r}$ can be easily obtained by directly computing $\mathcal{T}^{-1}(\mathbf{y})$. However,  $\mathcal{T}$ is often ill-posed, which means the inverse function $\mathcal{T}^{-1}$ does not exist and  the inverse problem of (\ref{for-inverse}) may have multiple solutions. Moreover, when the CT imaging is low-dose, the filter backward projection (FBP)~\cite{kak2001principles} can  produce serious detrimental artifact. Therefore, most of existing approaches usually incorporate some prior knowledge  during the reconstruction~\cite{li2020nett,lunz2018adversarial,rudin1992nonlinear}. For example, a commonly used method is based on 
%project $\mathbf{y}$ to  by the inverse of  forward operator. However, filter backward projection (FBP)~\cite{kak2001principles} can be applied in CT image reconstruction of standard-dose, which produce detrimental artifacts for low-dose CT.
%Since the inverse problem is ill-posedness without any prior knowledge,  several $\mathbf{x_g}$ can be recovered from $\mathbf{y}$.
%To overcome this challenge,  well-established approaches for reconstructing $\mathbf{x_g}$ from $\mathbf{y}$ 
%incorporate prior knowledge through 
regularization:
\begin{equation}
    \mathbf{x}= argmin_{\mathbf{x}} \left\| \mathcal{T}(\mathbf{x})-\mathbf{y}\right\|_p+\lambda \mathcal{R}(\mathbf{x}), \label{for-vr}
\end{equation}
where $\|\cdot\|_p$ denotes the $p$-norm and $\mathcal{R}(\mathbf{x})$ denotes the  penalty item from some prior knowledge. 

In the past years, a number of methods have been proposed for designing the regularization $\mathcal{R}$. 
%Various methods have been proposed to reduce the artifacts and improve the quality of CT image reconstruction. 
The traditional model-based algorithms, e.g., the ones using total variation~\cite{chambolle2004algorithm,rudin1992nonlinear}, usually apply the  sparse gradient assumptions and run an iterative algorithm   to learn the regularizers~\cite{knoll2011second,mccann2016fast,romano2017little,venkatakrishnan2013plug}. 
Another popular line for learning the regularizers  comes from deep learning 
%Alternatively, deep learning methods have been extensively explored for learning the regularizers  
\cite{kobler2020total,lunz2018adversarial}; the advantage of the deep learning methods is that they can achieve an end-to-end recovery of the true image $\mathbf{x^r}$  from the measurement $\mathbf{y}$~\cite{adler2018learned,mukherjee2021end}. Recent researches reveal that convolutional neural networks (CNNs) are quite effective for image denoising, e.g., the CNN based algorithms~\cite{jin2017deep,wolterink2017generative} can directly learn the reconstructed mapping from initial measurement reconstructions (e.g., FBP) to the ground-truth images.
%Wolterink \textit{et.al}~\cite{wolterink2017generative} took advantage of generative adversarial networks (GANs) to reduce noise, where generator mapped low-dose CT to standard-dose. 
The dual-domain network  that combines the sinograms with reconstructed low-dose CT images was also proposed to enhance the generalizability~\cite{lin2019dudonet,wang2022dudotrans}.

A major drawback of the aforementioned reconstruction methods is that they deal with the input CT 2D slices independently (note that the goal of CT reconstruction is to build the 3D model of the organ). 
%When reconstructing low-dose CT images using inverse problem algorithms, whether through model-based or data-driven methods based on deep learning, multiple CT images of the human body are typically treated as independent data items.
Namely, the neighborhood correlations among the 2D slices are often ignored, which may affect the reconstruction performance in practice. 
%There is no considering about any correlation between the images of same individual. 
In the field of computer vision, ``optical flow'' is a common technique  for tracking the motion of object between consecutive frames, which has been applied to many different tasks like  video generation~\cite{xue2016visual}, prediction of next frames~\cite{patraucean2015spatio} and  super resolution synthesis~\cite{chu2020learning,wang2018video}. To estimate the optical flow field, exsting approaches include the  traditional brightness gradient methods~\cite{beauchemin1995computation} and the deep networks \cite {dosovitskiy2015flownet}. The idea of optical flow has also been used for tracking  the organs movement in medical imaging~\cite{liu2020low,mira20203d,weng1997three}. However, to the best of our knowledge, there is no work considering  GANs with using optical flow  to capture neighbor slices coherence  for low dose 3D CT reconstruction. 

%Additionally,  3D Gans for reconstruction~\cite{kudo2019virtual} resulted in expensive computational overhead and  artifacts in single slice.
In this paper, we propose a novel optical flow based generative adversarial network for 3D CT reconstruction. Our intuition is as follows. 
When a patient is located in a CT equipment, a set of consecutive  cross-sectional images  are generated. If the vertical axial sampling space of transverse planes is small, 
%the images of the interior tissue  depicted by 
the corresponding CT slices should be  highly similar. 
%, resulting in close brightness for adjacent images. 
%Due to significant effect in tracking the transfer of the target, 
So we apply optical flow, though there exist several technical issues waiting to solve for the design and implementation, to capture the local coherence of adjacent CT images for reducing the artifacts in low-dose CT reconstruction. 
%it can also improve the image clarity of low-dose CT reconstruction through adjacent CT images. 
Our contributions are summarized below:
\begin{enumerate}
\item We introduce the ``local coherence'' by characterizing the  correlation of  consecutive CT   images, which plays a key role for 
%which takes full advantage of the smoothness between sequential images for  
suppressing the artifacts. 

\item Together with the local coherence, our proposed 
generative adversarial networks (GANs)  can yield significant improvement for texture quality and stability of the reconstructed images.

\item To illustrate the efficiency of our proposed approach, we conduct rigorous experiments on several real clinical datasets; the experimental results reveal the advantages of our approach over several state-of-the-art CT reconstruction methods. 
%The efficacy of the proposed approach is demonstrated through rigorous experiments on real clinical datasets, showcasing superior performance compared to state-of-the-art reconstruction methods.
\end{enumerate}
%two items how to merge!!!

\section{Preliminaries} %section name!!!
In this section, we briefly review the framework of the ordinary generative adversarial network,  and also introduce the local coherence of CT slices.
%that can be evaluated by optical flow.  

\subsubsection{Generative adversarial network.}
Traditional generative adversarial network~\cite{GoodfellowPMXWOCB14} consists of two main modules, a generator and a discriminator. The generator $\mathcal{G}$ is a mapping from a latent-space Gaussian distribution $\mathbb{P}_Z$ to the synthetic sample distribution $\mathbb{P}_{X_G}$, which is expected to be close to the real sample distribution $\mathbb{P}_X$. 
On the other hand,  the discriminator $\mathcal{D}$ aims to maximize the distance between the  distributions $\mathbb{P}_{X_G}$ and $\mathbb{P}_X$. 
%The discriminator $\mathcal{D}$ is essentially a classifier that identifies whether the sample comes from  real sample distribution $\mathbb{P}_X$ (judgment output value is $1$) or fake distribution from the generator $\mathbb{P}_{X_G}$ (judgment output value is $0$). 
%This results in an adversarial process, which can be seen as a game between the generator and discriminator. 
The game between the generator and discriminator actually is an adversarial process, where the overall optimization objective follows a min-max principle:
\begin{equation} 
    \mathop{min}\limits_{\mathcal{G}} \mathop{max}\limits_{\mathcal{D}} \mathbb{E}_{\mathbf{x^r} \sim \mathbb{P}_X, \mathbf{z} \sim \mathbb{P}_Z}(\log(\mathcal{D}(\mathbf{x^r}))+\log(1-\mathcal{D}(\mathcal{G}(\mathbf{z}))).
\end{equation}
%The generator and discriminator are optimized in turn, while another network is fixed. After many iterations of optimization, the two networks finally reach the state of Nash equilibrium. 

%\subsubsection{Local coherence captured by optical flow.}
%
\subsubsection{Local coherence.}
As mentioned in Section~\ref{introduction}, optical flow can capture the temporal coherence of object movements,  which plays a crucial role in many video-related tasks. More specifically, the optical flow refers to the instantaneous velocity of pixels of moving objects on consecutive frames over a short period of time~\cite{beauchemin1995computation}. The main idea relies on the practical assumptions that the brightness of the object more likely remains stable across consecutive frames, and the brightness of the pixels in a local region are usually changed consistently~\cite{horn1981determining}.
%synchronized in a local region
Based on these assumptions, the brightness of optical flow can be described by the following equation:
\begin{equation}\label{eqa:optical flow}
    \nabla I_w \cdot v_w + \nabla I_h \cdot v_h + \nabla I_t=0,
\end{equation} 
where  $v=(v_w,v_h)$ represents the optical flow of the position $(w,h)$ in the image. $\nabla I =(\nabla I_w,\nabla I_h) $ denotes spatial gradients of image brightness, and $\nabla I_t$ denotes the temporal partial derivative of the corresponding region.  

\begin{figure}[htbp]
    \centering
    \includegraphics[width=0.7\columnwidth]{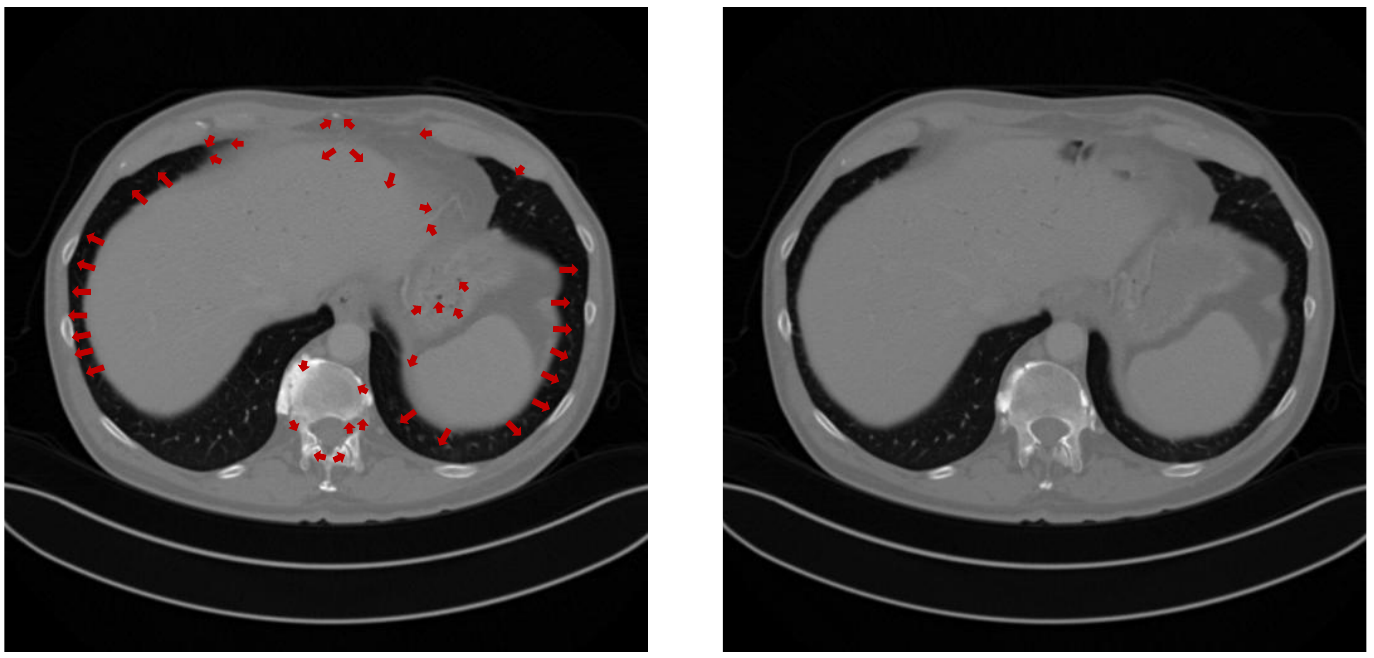}
    \caption{The optical flow between two adjacent CT slices.
    The scanning window of X-ray slides from the position of the left image to the position of the right image. The directions and lengths of the red arrows represent the optical flow field. The left and right images share the local coherence and thus the optical flows are small. }
    \label{fig:flow}
\end{figure}

Following the equation~(\ref{eqa:optical flow}), we consider the question that whether the optical flow idea can be applied to 3D CT reconstruction.
%for their high similarity in local slices.
%As a common knowledge, a CT scan of the human body is comprised of a series of sequential 2D images. 
In practice, the brightness of adjacent CT images often has very tiny difference, due to the inherent continuity and structural integrity of human body.
%From the perspective of tracking the target trajectory, such differences between sequential CT images represent the micro deformation of the  tissues across transverse plane, which can be captured by optical flow. 
Therefore, we introduce the ``local coherence'' that indicates the correlation  between adjacent images of a tissue.  
%%%%%%%%%%%%%%%%%%%%%%%%%
Namely, 
%the content of 
adjacent CT images often exhibit significant similarities within a certain local range along the vertical axis of the human body. 
%, which is referred to as local coherence. 
Due to the  local coherence, the noticeable  variations observed in CT slices within the local range often occur at  the edges of organs. 
%%%%%%%%%%%%%%%%%%%%%%%%%%
We can  substitute the temporal partial derivative $\nabla I_t$ by the  vertical axial partial derivative  $\nabla I_z$ in the equation~(\ref{eqa:optical flow}), where ``z'' indicates the index  of the  vertical axis. 
As illustrated in Figures \ref{fig:flow}, the local coherence can be captured by the optical flow  between adjacent CT slices.   
%To meet the high precision requirements of CT scans, we impose additional constraints of optical flow during the reconstruction process, which is more concrete in next section.

\section{GANs with local coherence}
In this section, we introduce our  low-dose CT image generation framework with local coherence in detail.  
%\href{https://github.com/lwjie595/GANLC}{https://github.com/lwjie595/GANLC}.
%Our framework contains four modules: framework of our network, optical flow estimator, generator and discriminator.

\subsubsection{The framework of our network.}
The proposed framework comprises three components, including a generator $\mathcal{G}$, a discriminator $\mathcal{D}$ and an optical flow estimator $\mathcal{F}$. The generator is the core component, and the flow estimator provides auxiliary warping images for the generation process.

\begin{figure}[htbp]
    \centering
    \includegraphics[width=1.0\columnwidth]{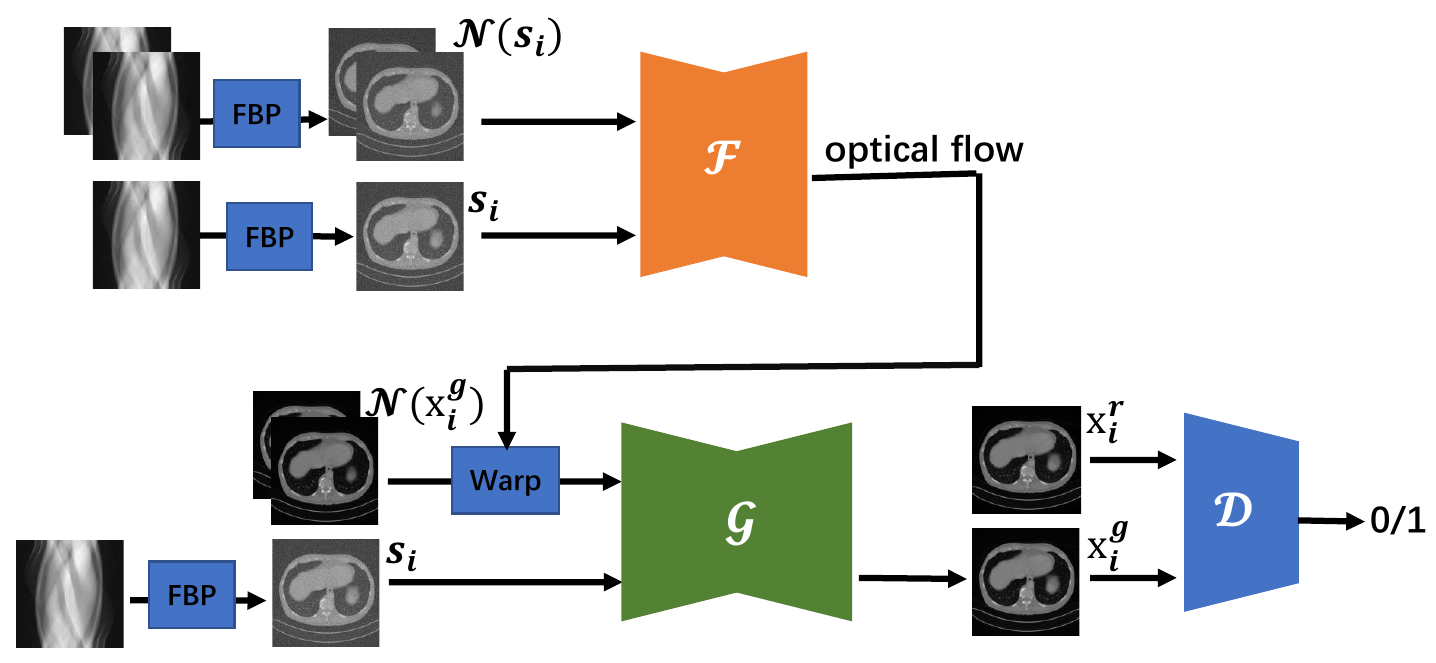}
    \caption{The framework of our generate adversarial network with local coherence  for CT reconstruction.}
    \label{fig:network}
\end{figure}

Suppose we have a sequence of measurements $\mathbf{y_1}, \mathbf{y_2}, \cdots,\mathbf{ y_n}$; for each $\mathbf{y_i}$, $1\leq i \leq n$, we want to reconstruct its ground truth image $\mathbf{x_i^r}$ as the equation~(\ref{for-inverse}).
Before performing the reconstruction in the generator $\mathcal{G}$,
we apply some  prior knowledge in physics and run filter backward projection on the measurement $\mathbf{y_i}$ in equation~(\ref{for-inverse})  to obtain an initial recovery solution $\mathbf{s_i}$. Usually $\mathbf{s_i}$ contains significant noise comparing with the ground truth $\mathbf{x_i^r}$.
Then the network has two input components, i.e., the initial backward projected image $\mathbf{s_i}$ that serves as an approximation of the ground truth $\mathbf{x_i^r}$, and  a set of neighbor CT slices $\mathcal{N}(\mathbf{s_i})=\{\mathbf{s_{i-1}},\mathbf{s_{i+1}}\}$ \footnote{If $i=1$,  $\mathcal{N}(\mathbf{s_i})=\{\mathbf{s_{2}}\}$; if $i=n$, $\mathcal{N}(\mathbf{s_i})=\{\mathbf{s_{n-1}}\}$. } for preserving the local coherence.
%which is important for preserving local coherence. 
%The main process is that $\mathcal{F}$ generates the optical flow from these inputs, which is subsequently used in conjunction with $\mathcal{N}(\mathbf{x_g})$ to perform a warping operation to obtain the corresponding reconstructed image $\mathbf{x_g}$.
%The proposed framework demonstrates promising results in the reconstruction of standard-dose CT images.
%The network is a  post-processing  technique that consider  correlation between adjacent slices in the reconstruction process, without utilizing filter function and Radon inverse projection for measurements. 
The overall structure of our framework is shown in Figure \ref{fig:network}. Below, we introduce the three key parts of our framework separately. 
%The images $\mathbf{s}$ and $\mathcal{N}(\mathbf{s})$ are put into $\mathcal{F}$  to generate optical flow. Then, $\mathbf{s}$ is combined with the image $\mathcal{W}(\mathcal{N}(\mathbf{x_g}))$, which generated by $\mathcal{N}(\mathbf{s})$ through $\mathcal{G}$ with warping operation, to obtain the corresponding reconstructed image $\mathbf{x_g}$ through the generator. Here, warping operation $\mathcal{W}$ performing on $\mathcal{N}(\mathbf{x_g})$ is based on the optical flow  between $\mathcal{N}(\mathbf{s})$ and $\mathbf{s}$.

\subsubsection{Optical flow estimator.}
%We have explained that optical flow can evaluate the local coherence between adjacent CT slices previously. 
%As we know, the CT images from one body are a series of cross section, so, for two adjacent initial CT slice images $\mathbf{s'}$ and $\mathbf{s}$, 
The optical flow $\mathcal{F}(\mathcal{N}(\mathbf{s_i}),\mathbf{s_i})$ denotes the brightness changes of  pixels from    $\mathcal{N}(\mathbf{s_i})$ to $\mathbf{s_i}$, where it captures their local coherence. 
%image $s$ and images  $\mathcal{N}(\mathbf{s_i})$. 
The estimator is derived by the network architecture of FlowNet~\cite{dosovitskiy2015flownet}. The FlowNet is an autoencoder architecture with extraction of features of two input frames to learn the corresponding flow,
%%%%%%%%%%%%%%%%%%
which is consist of $6$ (de)convolutional layers for both encoder and decoder.  %%%%%%%%%%%%%%%%%%%%%%%
% The learning objective of the optical flow estimator is:
% \begin{equation}\label{Flow_loss}
%   \mathcal{L}_{\mathcal{F}}=\sum_{\mathbf{s}} \|\mathcal{W}(\mathbf{s'},\mathcal{F}(\mathbf{s},\mathbf{s'}))-\mathbf{s}\|
% \end{equation}
%The $\mathcal{L}_{\mathcal{F}}$ is the discrepancy of pixel-wise between the warped image $\mathcal{W}(\mathbf{s'},\mathcal{F}(\mathbf{s},\mathbf{s'}))$ and image $\mathbf{s}$.
%After that, we  warp images in $\mathcal{N}(\mathbf{s_i})$ towards $\mathbf{s_i}$, according to the flow $\mathcal{F}(\mathcal{N}(\mathbf{s_i}),\mathbf{s_i})$. The warping operator $\mathcal{W} $ utilizes bi-linear interpolation.
%By leveraging the warped CT image, the authors demonstrate the ability to supply reference and address deficiencies of information in CT image s that contains organ or tissue areas with blurry parts. This is achieved by interpolating pixels of the warped image based on the pixels in the adjacent images N(s). Overall, the proposed approach offers a promising solution for improving the quality of CT images and enhancing the accuracy of medical image analysis.
%By leveraging the warped CT images, we demonstrate the ability of local coherence to address the deficiencies of information for  CT image $\mathbf{s_i}$ with blurry parts. This is achieved by interpolating pixels of the adjacent images $\mathcal{N}(\mathbf{s_i})$ to obtain warped images.

\subsubsection{Discriminator.}
The discriminator $\mathcal{D}$ assigns the label ``$1$'' to real  standard-dose CT images and ``$0$'' to generated images. The goal of  $\mathcal{D}$  is to maximize the separation between the distributions of real images and generated images:
\begin{equation} 
        \mathcal{L}_{\mathcal{D}}=\sum_{i=1}^n - (\log(\mathcal{D}(\mathbf{x_i^r}))+\log(1-\mathcal{D}(\mathbf{x_i^g}))),
\end{equation}
where $\mathbf{x_i^g}$ is the  image generated by  $\mathcal{G}$ (the formal definition for $\mathbf{x_i^g}$ will be introduced below).
%%%%%%%%%%%%%%%%%%%%%%%%%
The discriminator includes 3 residual blocks, with 4 convolutional layers in each residual block.
%%%%%%%%%%%%%%%%%%%%%%%%%
  
\subsubsection{Generator.}
%Different with conventional GAN architecture, the proposed generator takes into account the $\mathcal{N}(\mathbf{s})$ and $\mathbf{s}$ as inputs.
%where $\mathcal{N}(\mathbf{s})$ represents the neighbor slices set of $\mathbf{s}$.
We use the generator $\mathcal{G}$ to reconstruct the high-quality CT image for the ground truth $\mathbf{x_i^r}$   from the low-dose image $\mathbf{s_i}$. 
%The warping operator $\mathcal{W}(\cdot) $ is a bi-linear interpolation 
%The high-level structure of generator can be conceptualized as:
The generated image is obtained by

\begin{equation} \label{eqa:generator}
\begin{aligned}
    &\mathbf{x_i^g}= \mathcal{G}(\mathbf{s_i},\mathcal{W}(\mathcal{N}(\mathbf{x_i^g})));\\
    & \mathcal{N}(\mathbf{x_i^g})=\mathcal{G}(\mathcal{N}(\mathbf{s_i})), \\
\end{aligned}
\end{equation} 
%\text{with} ~\mathcal{N}(\mathbf{x_i^g})=\mathcal{G}(\mathcal{N}(\mathbf{s_i}))
where $\mathcal{W}(\cdot)$ is the warping operator.
Before generating $\mathbf{x_i^g}$, $\mathcal{N}(\mathbf{x_i^g})$ is reconstructed from $\mathcal{N}(\mathbf{s_i})$ by the generator without considering local coherence. Subsequently, according to the optical flow $\mathcal{F}(\mathcal{N}(\mathbf{s_i}),\mathbf{s_i})$, we warp the reconstructed images $\mathcal{N}(\mathbf{x_i^g})$ to align with the current slice by adjusting the brightness values. The warping operator $\mathcal{W} $ utilizes bi-linear interpolation to obtain $\mathcal{W}(\mathcal{N}(\mathbf{x_i^g}))$, which enables the model to capture subtle variations in the tissue from the generated $\mathcal{N}(\mathbf{x_i^g})$;  also, the warping operator can reduce the influence  of artifacts for the reconstruction. Finally,  $\mathbf{x_i^g}$ is generated by combining $\mathbf{s_i}$ and $\mathcal{W}(\mathcal{N}(\mathbf{x_i^g}))$.  Since $\mathbf{x_i^r}$ is our target for reconstruction in the $i$-th batch, we consider the difference between $\mathbf{x_i^g}$ and $\mathbf{x_i^r}$ in the loss. 
%whereas images in $\mathcal{N}(\mathbf{x_g})$ serve as auxiliary compensation information to address the blurriness issues in the reconstruction of $\mathbf{x_g}$.
Our generator is mainly based on the network  architecture of Unet~\cite{ronneberger2015u}.
Partly inspired by the loss in \cite{armanious2020medgan}, the optimization objective of the generator $\mathcal{G}$ comprises three items with the coefficients $ \lambda_{\mathtt{pix}},\lambda_{\mathtt{adv}},\lambda_{\mathtt{per}} \in (0,1)$: %$\mathcal{L}_{content}$, $\mathcal{L}_{adv}$ and $\mathcal{L}_{percept}$, which can be summarized as:
\begin{equation} \label{eqa:loss}
    \mathcal{L}_{\mathcal{G}}=\lambda_{\mathtt{pix}} \mathcal{L}_{\mathtt{pixel}} + \lambda_{\mathtt{adv}} \mathcal{L}_{\mathtt{adv}} +\lambda_{\mathtt{per}} \mathcal{L}_{\mathtt{percept}}.
\end{equation}
In (\ref{eqa:loss}), ``$\mathcal{L}_\mathtt{pixel}$'' is the loss measuring the pixel-wise mean square error of the generated image $\mathbf{x_i^g}$ with respect to the ground-truth $\mathbf{x_i^r}$. 
``$\mathcal{L}_\mathtt{adv}$'' represents the adversarial loss of the discriminator $\mathcal{D}$, which is designed to minimize the distance between the generated standard-dose CT image distribution $\mathbb{P}_{X_G}$ and the real standard-dose CT image distribution $\mathbb{P}_{X}$. 
%(the formal definition for $\mathcal{D}$ will be introduced below). 
``$\mathcal{L}_\mathtt{percept}$'' denotes the perceptual loss, which quantifies the dissimilarity between the feature maps of $\mathbf{x_i^r}$ and  $\mathbf{x_i^g}$; the feature maps denote the feature representation extracted from the hidden layers in the discriminator $\mathcal{D}$ (suppose there are $t$ hidden layers):
\begin{equation}\label{Flow_loss}
   \mathcal{L}_{percept}=\sum_{i=1}^n \sum_{j=1}^t  \|\mathcal{D}_j(\mathbf{x_i^r})-\mathcal{D}_j(\mathbf{x_i^g})\|_1
\end{equation}
where $\mathcal{D}_j(\cdot)$ refers to the feature extraction performed on the $j$-th hidden layer. Through capturing  the high frequency differences in CT images, $\mathcal{L}_\mathtt{percept}$ can enhance the sharpness for edges and increase the contrast for the reconstructed images. $\mathcal{L}_\mathtt{pixel}$ and $\mathcal{L}_\mathtt{adv}$ are designed to recover global structure, 
%which are based on intuitive judgments of GANs and image pixels, 
and $\mathcal{L}_\mathtt{percept}$ is utilized to incorporate additional texture details into the reconstruction process.
%%%%%%%%%%%%%%%%%%%%%%%%%

%%%%%%%%%%%%%%%%%%%%%%%%%%

\section{Experiment}
\subsubsection{Datasets.}
First, our proposed approaches are evaluated on the ``Mayo-Clinic low-dose CT Grand Challenge"  (Mayo-Clinic) dataset of lung CT images~\cite{mccollough2016tu}. The dataset contains 2250 two dimensional slices from 9 patients for training, and the remaining 128 slices from 1 patient are reserved for testing. The low-dose measurements are simulated by parallel-beam X-ray  with 200 (or 150) uniform  views, i.e., $N_v=200 ~(\text{or}~ N_v=150)$, and 400 (or 300) detectors, i.e., $N_d=400 ~(\text{or}~ N_d=300)$. In order to further verify the denoising ability of our approaches, we add the Gaussian noise with standard deviation $\sigma=2.0$ to the sinograms after X-ray projection in 50\% of the experiments. 
To evaluate the generalization of our model, we also consider another dataset  RIDER  with non-small cell lung cancer under two CT scans~\cite{zhao2009evaluating} for testing. We randomly select  4 patients with 1827 slices from the dataset. 
%For this dataset, 4 patients with 1827 slices are selected for testing and 
The simulation process is identical to that of Mayo-Clinic. 
The proposed networks were implemented in the MindSpore framework and trained on Nvidia 3090 GPU with 100 epochs. 

\subsubsection{Baselines and evaluation metrics.}
We consider several existing popular algorithms for comparison.
%including existing mainstream deep learning and classical algorithms.
(1) \textbf{FBP}~\cite{kak2001principles}: the classical filter backward projection on low-dose sinograms. (2) \textbf{FBPConvNet}~\cite{jin2017deep}:  a direct inversion network followed by the CNN after initial \textbf{FBP} reconstruction. (3) \textbf{LPD}~\cite{adler2018learned}: a deep learning method  based on proximal primal-dual optimization.  (4) \textbf{UAR}~\cite{mukherjee2021end}: an end-to-end reconstruction method based on learning unrolled reconstruction operators and adversarial regularizers. 
Our proposed method is denoted by \textbf{GAN-LC}. 
%Due to the reference to the common hyperparameter setting, 
We set $ \lambda_{\mathtt{pix}}=1.0,\lambda_{\mathtt{adv}}=0.01~ \text{and }~\lambda_{\mathtt{per}}=1.0 $  for the optimization objective in equation~(\ref{eqa:loss}) during our training process.
Following most of the previous articles on 3D CT reconstruction, we evaluate the experimental performance by  two metrics: the peak signal-to-noise ratio (PSNR) and the structural similarity index (SSIM)~\cite{wang2004image}. PSNR measures the pixel differences of two images, which is negatively correlated with mean square error. 
%, higher value being better.
SSIM measures the structure similarity between two images, which is related to the  variances of the input images. For both two meansures, the higher the better. 

%, higher value being expected.

\subsubsection{Results.}
Table \ref{tab:mayor} presents the  results on the Mayo-Clinic dataset, where the first row represents different parameter settings (i.e., the number of uniform views $N_v$, the number of detectors $N_d$ and the standard deviation of Gaussian noise $\sigma$) for simulating low-dose sinograms. Our proposed approach  \textbf{GAN-LC} consistently outperforms the baselines under almost all the low-dose parameter settings.
%indicating its state-of-the-art performance. 
The methods \textbf{FBP} and \textbf{UAR} are very sensitive to noise; the performance of \textbf{LPD} is relatively stable but with low reconstruction accuracy. \textbf{FBPConvNet} has a similar increasing trend with our approach across different settings but has worse reconstruction quality.
To evaluate the stability and generalization of our model and the baselines trained on Mayo-Clinic dataset, we also test them on the RIDER dataset. 
%using the existing trained-well network from Mayo-Clinic dataset.
The results are shown in Table \ref{tab:rider}. Due to the bias in the datasets collected from different  facilities,  the performances of all the models are declined  to some extents.
But our proposed approach still outperforms the other models for most testing cases.

\begin{figure}[htbp]
    \centering
    \includegraphics[width=0.7\columnwidth]{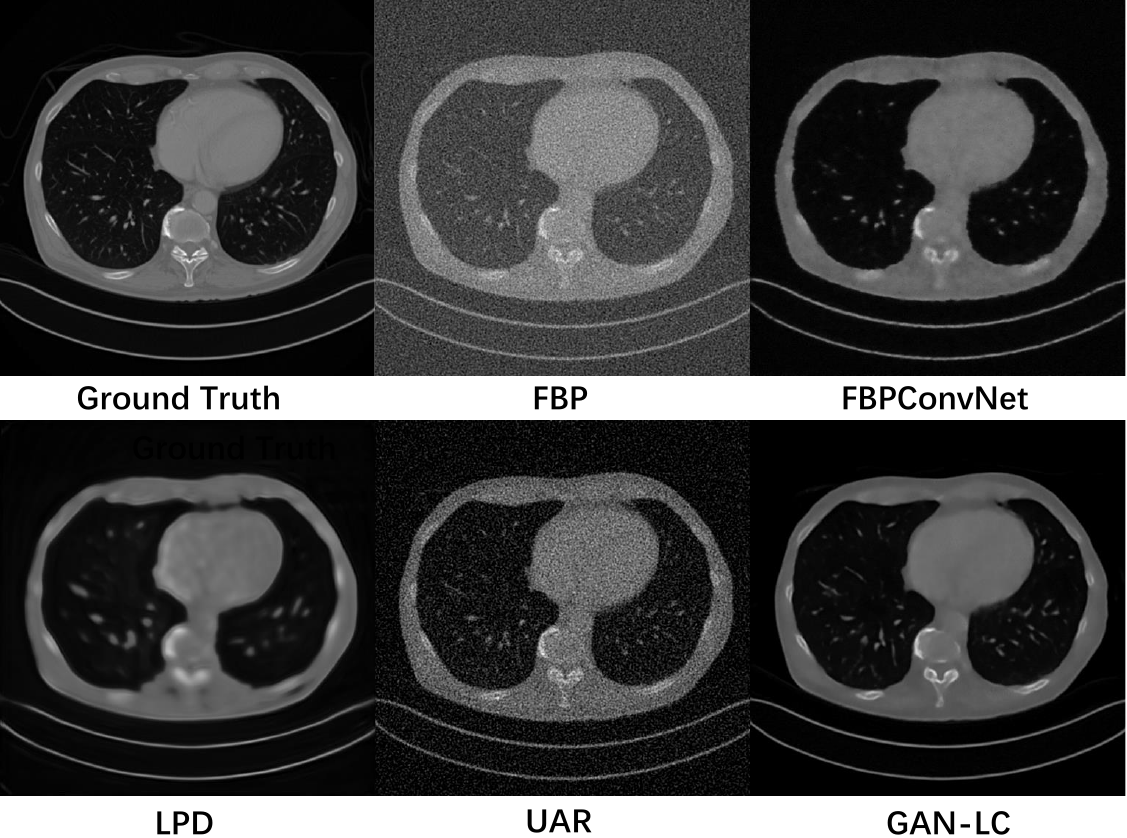}
    \caption{Reconstruction results on Mayo-Clinic dataset. The sparse view setting of sinograms  is $N_v=200$, $N_d=400$ and $\sigma=2.0$. ``Ground Truth'' is the standard-dose CT image.}
    \label{fig:result_mayor_noise}
\end{figure}

To illustrate the reconstruction performances more clearly, we also show the  reconstruction results for testing images in Figure \ref{fig:result_mayor_noise}. We can see that our network can reconstruct the CT image with higher quality. Due to the space limit, the experimental results of different views $N_v$ and more  visualized results  are placed in our supplementary material.

% \begin{figure}[htbp]
%     \centering
%     \includegraphics[width=0.52\columnwidth]{result2_1.pdf}
%     \includegraphics[width=0.52\columnwidth]{result3_1.pdf}
%     \caption{Reconstruction results on Mayo-Clinic dataset. The sparse view setting of sinograms  is $N_v=200$, $N_d=400$ and $\sigma=2.0$. Here, ``Ground Truth'' is the standard-dose CT image.}
%     \label{fig:result_mayor_noise}
% \end{figure}

\begin{table}[htbp]
  \centering
  \caption{Experimental results for Mayo-Clinic dataset. The value in first row of the table represents $N_v$, $N_d$ and $\sigma$  for simulating low-dose sinograms, respectively. }
    \begin{tabular}{c| c |c|c|c|c|c|c|c}
    \hline
  Sinograms  & \multicolumn{2}{ c |}{$~~200,400,0.0~~$} & \multicolumn{2}{c|}{$~~200,400,2.0~~$} & \multicolumn{2}{ c |}{$~~150,300,0.0~~$} & \multicolumn{2}{c}{$~~150,300,2.0~~$} \\ 	\hline
          &  PSNR  & SSIM  &  PSNR  & SSIM  &  PSNR  & SSIM  &  PSNR  &  SSIM  \\ 	\hline
    \textbf{FBP}   & 26.449 & 0.721 & 13.517 & 0.191 & 21.460 & 0.616 & 12.593 & 0.168 \\
    \textbf{FBPConvNet} & 38.213 & 0.918 & 30.148 & 0.743 & 35.263 & 0.869 & 29.095 & 0.723 \\
    \textbf{LPD}   & 28.050 & 0.844 & 28.357 & 0.794 & 28.376 & 0.826 & 27.409 & \textbf{0.801} \\
    \textbf{UAR}   & 33.248 & 0.902 & 22.048 & 0.272 & 29.829 & 0.848 & 21.227 & 0.238 \\
    \textbf{GAN-LC} & \textbf{39.548} & \textbf{0.950}  & \textbf{32.437} & \textbf{0.819} & \textbf{36.542} & \textbf{0.899} & \textbf{31.586} & 0.725 \\
    \hline
    \end{tabular}%
  \label{tab:mayor}%
\end{table}%

\begin{table}[htbp]
  \centering
  \caption{Experimental results for RIDER  dataset. The value in first row of the table represents $N_v$, $N_d$ and $\sigma$  for simulating low-dose sinograms, respectively.}
    \begin{tabular}{c| c |c|c|c|c|c|c|c}
         \hline
   Sinograms  & \multicolumn{2}{ c |}{$~~200,400,0.0~~$} & \multicolumn{2}{c|}{$~~200,400,2.0~~$} & \multicolumn{2}{ c |}{$~~150,300,0.0~~$} & \multicolumn{2}{c}{$~~150,300,2.0~~$} \\ 	\hline
          &  PSNR  & SSIM  &  PSNR  & SSIM  &  PSNR  & SSIM  &  PSNR  &  SSIM  \\ 	\hline
    \textbf{FBP}   & 21.398 & 0.647 & 15.609 & 0.233 & 19.49 & 0.597 & 14.845 & 0.203 \\
    \textbf{FBPConvNet} & 27.256 & 0.671 & 19.520 & 0.444 & 27.504 & 0.650  & 18.517 & 0.431 \\
    \textbf{LPD}   & 22.341 & 0.615 & 12.196 & 0.466 & 22.172 & 0.556 & 12.215 & 0.455 \\
    \textbf{UAR}   & 24.915 & 0.667 & 20.943 & 0.207 & 21.136 & 0.557 & \textbf{19.873} & 0.176 \\
    \textbf{GAN-LC} &\textbf{ 28.861} & \textbf{0.721} & \textbf{22.624} & \textbf{0.517} & \textbf{29.171} & \textbf{0.705} & 19.607 & \textbf{0.470} \\
    \hline
    \end{tabular}%
  \label{tab:rider}%
\end{table}%

\section{Conclusion}
In this paper, we propose a novel approach for low-dose CT reconstruction using generative adversarial networks  with local coherence. By considering the inherent continuity of human body, local coherence can be captured through optical flow, which is small deformations and structural differences between consecutive CT slices.
%, which can  make up for the prior knowledge in the spatial domain. 
%with denoising performance improved. 
The experimental results on real datasets demonstrate the advantages of our proposed network over several popular approaches.
%, and producing reconstruction results with details and textures that are closer to standard-dose CT images.
In  future, we will evaluate our network on real-world CT images from local hospital and use the reconstructed images to support doctors   for the diagnosis and recognition of  lung nodules. 
% Our code is publicly available at \url{https://github.com/lwjie595/GANLC}.
%\newlineincluding the identification and  of lung nodules and tumor tissues.
\newline
\newline
\textbf{Acknowledgements.} The research of this work was supported in part by National Key R\&D program of China through grant 2021YFA1000900, the NSFC throught Grant 62272432, the Provincial NSF of Anhui through grant 2208085MF163, a Huawei-USTC Joint Innovation Project on Fundamental System Software and sponsored by CAAI-Huawei MindSpore Open Fund. 
%Hu Ding is the corresponding author. 


\bibliographystyle{splncs04}
\bibliography{miccai.bib}

\begin{thebibliography}{10}
\providecommand{\url}[1]{\texttt{#1}}
\providecommand{\urlprefix}{URL }
\providecommand{\doi}[1]{https://doi.org/#1}

\bibitem{adler2018learned}
Adler, J., {\"O}ktem, O.: Learned primal-dual reconstruction. IEEE transactions
  on medical imaging  \textbf{37}(6),  1322--1332 (2018)

\bibitem{armanious2020medgan}
Armanious, K., Jiang, C., Fischer, M., K{\"u}stner, T., Hepp, T., Nikolaou, K.,
  Gatidis, S., Yang, B.: Medgan: Medical image translation using gans.
  Computerized medical imaging and graphics  \textbf{79},  101684 (2020)

\bibitem{beauchemin1995computation}
Beauchemin, S.S., Barron, J.L.: The computation of optical flow. ACM computing
  surveys (CSUR)  \textbf{27}(3),  433--466 (1995)

\bibitem{chambolle2004algorithm}
Chambolle, A.: An algorithm for total variation minimization and applications.
  Journal of Mathematical imaging and vision  \textbf{20}(1),  89--97 (2004)

\bibitem{chen2017low}
Chen, H., Zhang, Y., Zhang, W., Liao, P., Li, K., Zhou, J., Wang, G.: Low-dose
  ct via convolutional neural network. Biomedical optics express
  \textbf{8}(2),  679--694 (2017)

\bibitem{chu2020learning}
Chu, M., Xie, Y., Mayer, J., Leal-Taix{\'e}, L., Thuerey, N.: Learning temporal
  coherence via self-supervision for gan-based video generation. ACM
  Transactions on Graphics (TOG)  \textbf{39}(4),  75--1 (2020)

\bibitem{ding2021deep}
Ding, Q., Nan, Y., Gao, H., Ji, H.: Deep learning with adaptive
  hyper-parameters for low-dose ct image reconstruction. IEEE Transactions on
  Computational Imaging  \textbf{7},  648--660 (2021)

\bibitem{dosovitskiy2015flownet}
Dosovitskiy, A., Fischer, P., Ilg, E., Hausser, P., Hazirbas, C., Golkov, V.,
  Van Der~Smagt, P., Cremers, D., Brox, T.: Flownet: Learning optical flow with
  convolutional networks. In: Proceedings of the IEEE international conference
  on computer vision. pp. 2758--2766 (2015)

\bibitem{GoodfellowPMXWOCB14}
Goodfellow, I.J., Pouget{-}Abadie, J., Mirza, M., Xu, B., Warde{-}Farley, D.,
  Ozair, S., Courville, A.C., Bengio, Y.: Generative adversarial networks. CoRR
   \textbf{abs/1406.2661} (2014), \url{http://arxiv.org/abs/1406.2661}

\bibitem{horn1981determining}
Horn, B.K., Schunck, B.G.: Determining optical flow. Artificial intelligence
  \textbf{17}(1-3),  185--203 (1981)

\bibitem{jin2017deep}
Jin, K.H., McCann, M.T., Froustey, E., Unser, M.: Deep convolutional neural
  network for inverse problems in imaging. IEEE Transactions on Image
  Processing  \textbf{26}(9),  4509--4522 (2017)

\bibitem{kak2001principles}
Kak, A.C., Slaney, M.: Principles of computerized tomographic imaging. SIAM
  (2001)

\bibitem{knoll2011second}
Knoll, F., Bredies, K., Pock, T., Stollberger, R.: Second order total
  generalized variation (tgv) for mri. Magnetic resonance in medicine
  \textbf{65}(2),  480--491 (2011)

\bibitem{kobler2020total}
Kobler, E., Effland, A., Kunisch, K., Pock, T.: Total deep variation for linear
  inverse problems. In: Proceedings of the IEEE/CVF Conference on Computer
  Vision and Pattern Recognition. pp. 7549--7558 (2020)

\bibitem{li2020nett}
Li, H., Schwab, J., Antholzer, S., Haltmeier, M.: Nett: Solving inverse
  problems with deep neural networks. Inverse Problems  \textbf{36}(6),  065005
  (2020)

\bibitem{lin2019dudonet}
Lin, W.A., Liao, H., Peng, C., Sun, X., Zhang, J., Luo, J., Chellappa, R.,
  Zhou, S.K.: Dudonet: Dual domain network for ct metal artifact reduction. In:
  Proceedings of the IEEE/CVF Conference on Computer Vision and Pattern
  Recognition. pp. 10512--10521 (2019)

\bibitem{liu2020low}
Liu, H., Lin, Y., Ibragimov, B., Zhang, C.: Low dose 4d-ct super-resolution
  reconstruction via inter-plane motion estimation based on optical flow.
  Biomedical Signal Processing and Control  \textbf{62},  102085 (2020)

\bibitem{lunz2018adversarial}
Lunz, S., {\"O}ktem, O., Sch{\"o}nlieb, C.B.: Adversarial regularizers in
  inverse problems. Advances in neural information processing systems
  \textbf{31} (2018)

\bibitem{mccann2016fast}
McCann, M.T., Nilchian, M., Stampanoni, M., Unser, M.: Fast 3d reconstruction
  method for differential phase contrast x-ray ct. Optics express
  \textbf{24}(13),  14564--14581 (2016)

\bibitem{mccollough2016tu}
McCollough, C.: Tu-fg-207a-04: overview of the low dose ct grand challenge.
  Medical physics  \textbf{43}(6Part35),  3759--3760 (2016)

\bibitem{mira20203d}
Mira, C., Moya-Albor, E., Escalante-Ram{\'\i}rez, B., Olveres, J., Brieva, J.,
  Vallejo, E.: 3d hermite transform optical flow estimation in left ventricle
  ct sequences. Sensors  \textbf{20}(3), ~595 (2020)

\bibitem{mukherjee2021end}
Mukherjee, S., Carioni, M., {\"O}ktem, O., Sch{\"o}nlieb, C.B.: End-to-end
  reconstruction meets data-driven regularization for inverse problems.
  Advances in Neural Information Processing Systems  \textbf{34},  21413--21425
  (2021)

\bibitem{patraucean2015spatio}
Patraucean, V., Handa, A., Cipolla, R.: Spatio-temporal video autoencoder with
  differentiable memory. arXiv preprint arXiv:1511.06309  (2015)

\bibitem{ramm2020radon}
Ramm, A.G., Katsevich, A.I.: The Radon transform and local tomography. CRC
  press (2020)

\bibitem{romano2017little}
Romano, Y., Elad, M., Milanfar, P.: The little engine that could:
  Regularization by denoising (red). SIAM Journal on Imaging Sciences
  \textbf{10}(4),  1804--1844 (2017)

\bibitem{ronneberger2015u}
Ronneberger, O., Fischer, P., Brox, T.: U-net: Convolutional networks for
  biomedical image segmentation. In: Medical Image Computing and
  Computer-Assisted Intervention--MICCAI 2015: 18th International Conference,
  Munich, Germany, October 5-9, 2015, Proceedings, Part III 18. pp. 234--241.
  Springer (2015)

\bibitem{rudin1992nonlinear}
Rudin, L.I., Osher, S., Fatemi, E.: Nonlinear total variation based noise
  removal algorithms. Physica D: nonlinear phenomena  \textbf{60}(1-4),
  259--268 (1992)

\bibitem{sori2021dfd}
Sori, W.J., Feng, J., Godana, A.W., Liu, S., Gelmecha, D.J.: Dfd-net: lung
  cancer detection from denoised ct scan image using deep learning. Frontiers
  of Computer Science  \textbf{15},  1--13 (2021)

\bibitem{toft1996radon}
Toft, P.: The radon transform. Theory and Implementation (Ph. D.
  Dissertation)(Copenhagen: Technical University of Denmark)  (1996)

\bibitem{venkatakrishnan2013plug}
Venkatakrishnan, S.V., Bouman, C.A., Wohlberg, B.: Plug-and-play priors for
  model based reconstruction. In: 2013 IEEE Global Conference on Signal and
  Information Processing. pp. 945--948. IEEE (2013)

\bibitem{wang2022dudotrans}
Wang, C., Shang, K., Zhang, H., Li, Q., Zhou, S.K.: Dudotrans: Dual-domain
  transformer for sparse-view ct reconstruction. In: Machine Learning for
  Medical Image Reconstruction: 5th International Workshop, MLMIR 2022, Held in
  Conjunction with MICCAI 2022, Singapore, September 22, 2022, Proceedings. pp.
  84--94. Springer (2022)

\bibitem{wang2018video}
Wang, T.C., Liu, M.Y., Zhu, J.Y., Liu, G., Tao, A., Kautz, J., Catanzaro, B.:
  Video-to-video synthesis. arXiv preprint arXiv:1808.06601  (2018)

\bibitem{wang2004image}
Wang, Z., Bovik, A.C., Sheikh, H.R., Simoncelli, E.P.: Image quality
  assessment: from error visibility to structural similarity. IEEE transactions
  on image processing  \textbf{13}(4),  600--612 (2004)

\bibitem{weng1997three}
Weng, N., Yang, Y.H., Pierson, R.: Three-dimensional surface reconstruction
  using optical flow for medical imaging. IEEE transactions on medical imaging
  \textbf{16}(5),  630--641 (1997)

\bibitem{wolterink2017generative}
Wolterink, J.M., Leiner, T., Viergever, M.A., I{\v{s}}gum, I.: Generative
  adversarial networks for noise reduction in low-dose ct. IEEE transactions on
  medical imaging  \textbf{36}(12),  2536--2545 (2017)

\bibitem{xue2016visual}
Xue, T., Wu, J., Bouman, K., Freeman, B.: Visual dynamics: Probabilistic future
  frame synthesis via cross convolutional networks. Advances in neural
  information processing systems  \textbf{29} (2016)

\bibitem{zhao2009evaluating}
Zhao, B., James, L.P., Moskowitz, C.S., Guo, P., Ginsberg, M.S., Lefkowitz,
  R.A., Qin, Y., Riely, G.J., Kris, M.G., Schwartz, L.H.: Evaluating
  variability in tumor measurements from same-day repeat ct scans of patients
  with non--small cell lung cancer. Radiology  \textbf{252}(1),  263--272
  (2009)

\end{thebibliography}
\newpage
\appendix
\section{Supplementary figures}
\begin{figure*}[htbp]
	\centering
	{\includegraphics[width=0.455\columnwidth]{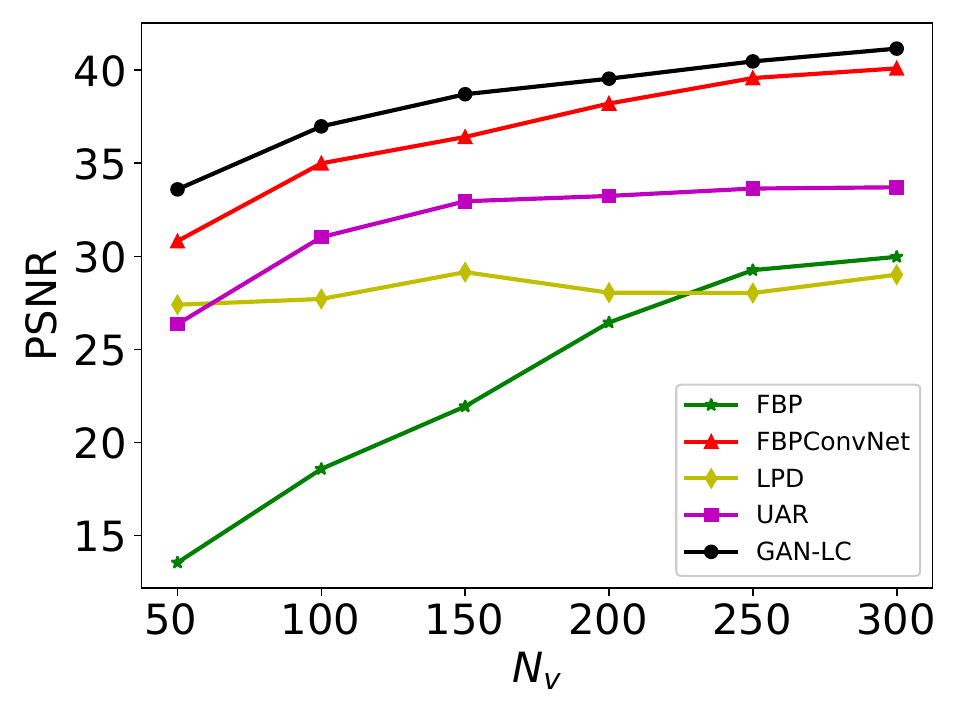} %[图片大小]{图片路径}
		\includegraphics[width=0.455\columnwidth]{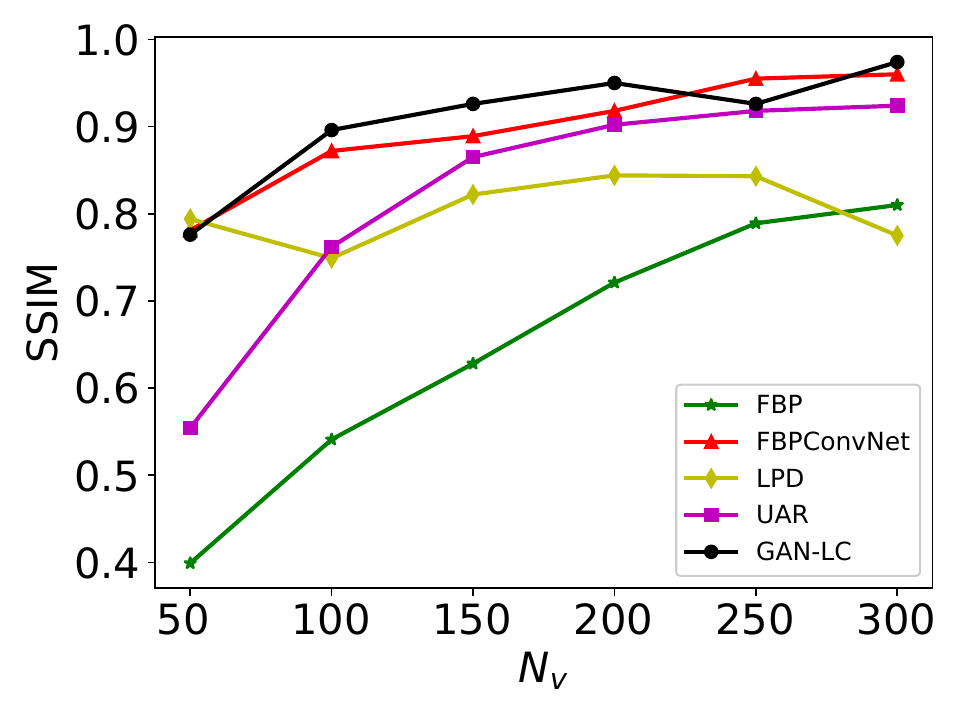}}
	\caption{Reconstruction results on Mayo-Clinic dataset with $N_v$  varying from $50$ to $300$.  The sparse view setting of sinograms is $N_d=400$ and $\sigma=0$. Our proposed approach \textbf{GAN-LC}  outperforms the baselines under almost all the low-dose parameter settings.} %图片标题
	\label{fig:mayor_vary_views}  %图片交叉引用时的标签
\end{figure*}
\begin{figure}[htbp]
	\centering
	\includegraphics[width=0.67\columnwidth]{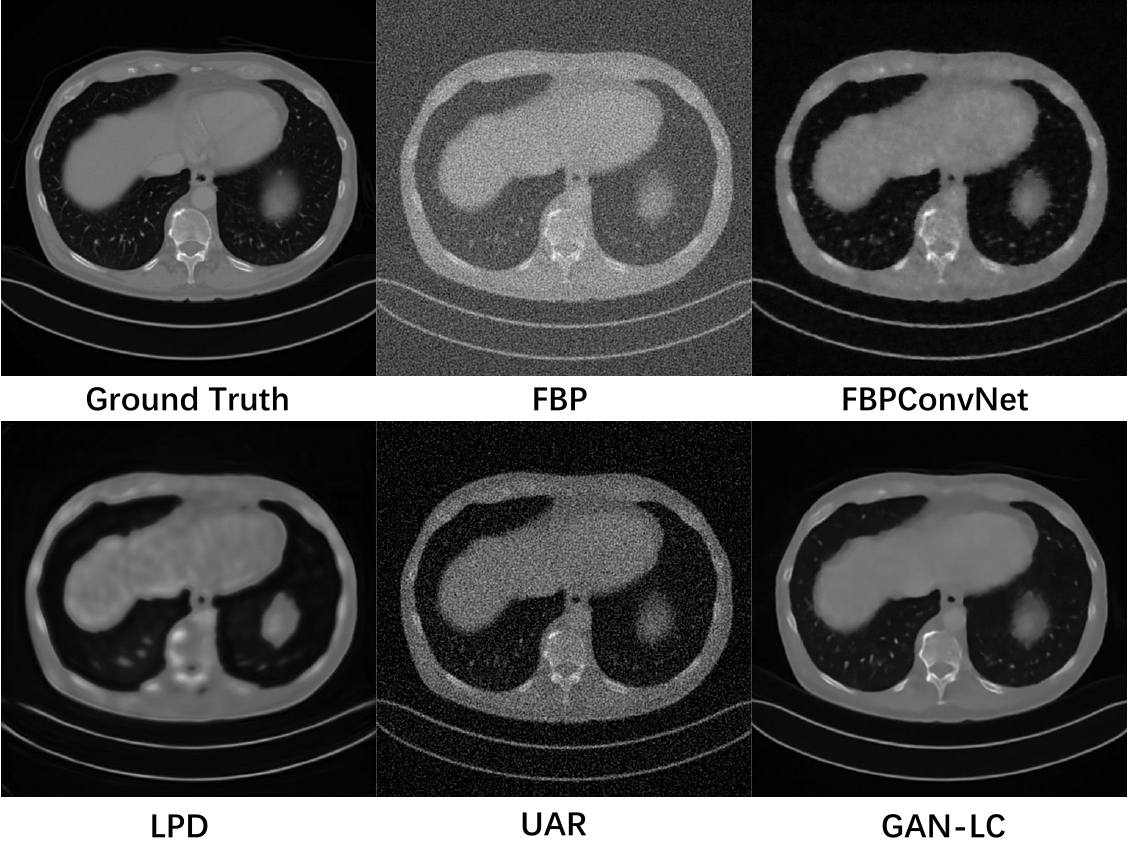}
	\caption{Reconstruction results on Mayo-Clinic dataset. The sparse view setting of sinograms  is $N_v=200$, $N_d=400$ and $\sigma=2.0$.}
	\label{fig:result_mayor_noise2}
\end{figure}

\begin{figure}[htbp]
	\centering
	\includegraphics[width=0.67\columnwidth]{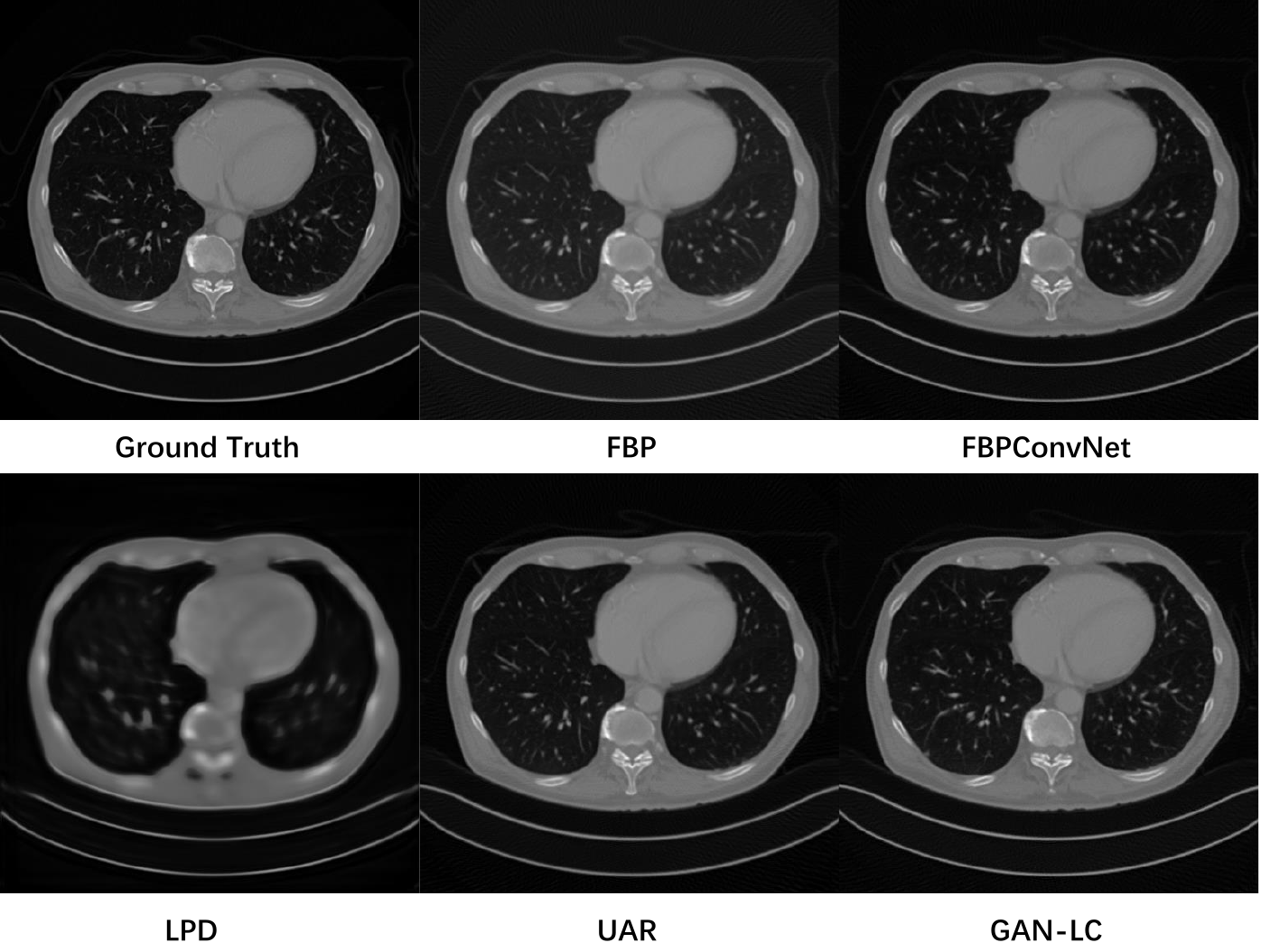}
	\caption{Reconstruction results on Mayo-Clinic dataset.  The sparse view setting of sinograms  is $N_v=200$, $N_d=400$ and $\sigma=0$.  }
	\label{fig:result_mayor_nonoise}
\end{figure}
% \begin{figure}[htbp]
	%     \centering
	%     \includegraphics[width=0.8\columnwidth]{rider_res1.pdf}
	%     \caption{Reconstruction results on Rider dataset. The sparse view setting of sinograms  is $N_v=200$, $N_d=400$ and $\sigma=0$.}
	%     \label{fig:result_rider_nonoise}
	% \end{figure}
\begin{figure}[htbp]
	\centering
	\includegraphics[width=0.67\columnwidth]{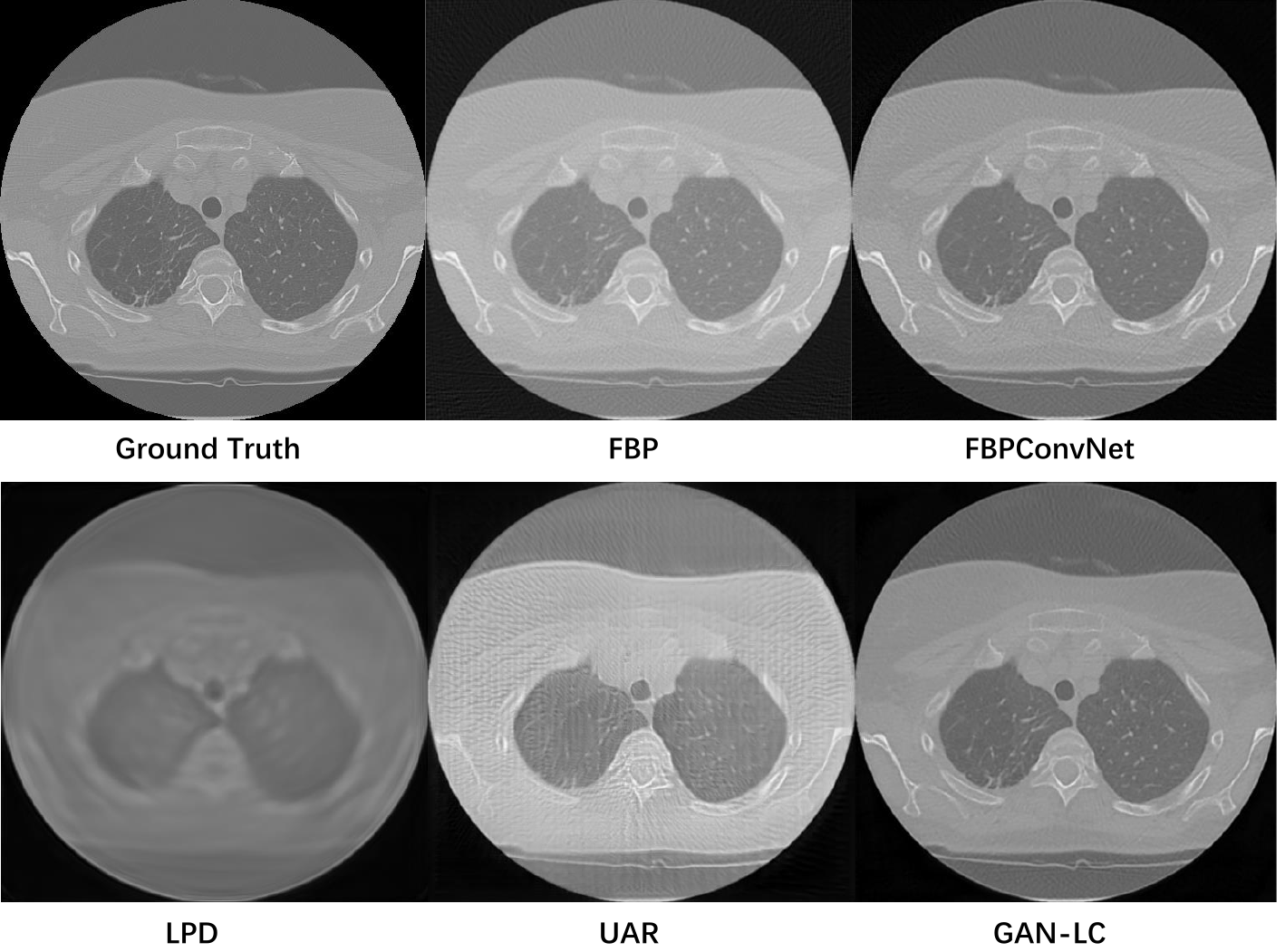}
	\caption{Reconstruction results on Rider dataset. The sparse view setting of sinograms  is $N_v=200$, $N_d=400$ and $\sigma=2.0$.}
	\label{fig:result_rider_noise}
\end{figure}

\end{document}